\documentclass[12pt]{spieman}  
\usepackage{amsmath,amsfonts,amssymb}
\usepackage{subcaption}
\usepackage{graphicx}
\usepackage{setspace}
\usepackage{tocloft}
\usepackage{booktabs}
\usepackage[pagewise]{lineno}

\title{Spectral Unmixing of Hyperspectral Images Based on Block Sparse Structure}

\author[a]{Seyed Hossein Mosavi Azarang}
\author[a,*]{Roozbeh Rajabi}
\author[a]{Hadi Zayyani}
\author[b]{Amin Zehtabian}
\affil[a]{Qom University of Technology, Faculty of Electrical and Computer Engineering, Department of Communications and Electronics, Qom, Iran, 3718146645}
\affil[b]{Freie Universität Berlin, Institut für Biochemie, Berlin, Germany, 14195}

\cftpagenumbersoff{figure}
\cftpagenumbersoff{table} 
\begin{document} 
\maketitle

\begin{abstract}
Spectral unmixing (SU) of hyperspectral images (HSIs) is one of the important areas in remote sensing (RS) that needs to be carefully addressed in different RS applications. Despite the high spectral resolution of the hyperspectral data, the relatively low spatial resolution of the sensors may lead to mixture of different pure materials within the image pixels. In this case, the spectrum of a given pixel recorded by the sensor can be a combination of multiple spectra each belonging to a unique material in that pixel. Spectral unmixing is then used as a technique to extract the spectral characteristics of the different materials within the mixed pixels and to recover the spectrum of each pure spectral signature, called endmember. Block-sparsity exists in hyperspectral images as a result of spectral similarity between neighboring pixels. In block-sparse signals, the nonzero samples occur in clusters and the pattern of the clusters is often supposed to be unavailable as prior information. This paper presents an innovative spectral unmixing approach for HSIs based on block-sparse structure. Hyperspectral unmixing problem is solved using pattern coupled sparse Bayesian learning strategy (PCSBL). To evaluate the performance of the proposed SU algorithm, it is tested on both synthetic and real hyperspectral data and the quantitative results are compared to those of other state-of-the-art methods in terms of abundance angle distance and mean squared error. The achieved results show the superiority of the proposed algorithm over the other competing methods by a significant margin.
\end{abstract}


{\noindent \footnotesize\textbf{*}Roozbeh Rajabi,  \linkable{rajabi@qut.ac.ir} }

\begin{spacing}{2}   

\section{Introduction}
\label{sec:intro}  
Remote sensing is the science of collecting and interpreting information from objects without physical contact with the scene. One of the modern technologies in remote sensing is hyperspectral imaging. Hyperspectral sensors have the capability of recording electromagnetic energy reflected on the surface of objects in very narrow spectral bands \cite{landgrebe2003signal}.

\begin{figure}
	\centering
	\includegraphics[width=11cm]{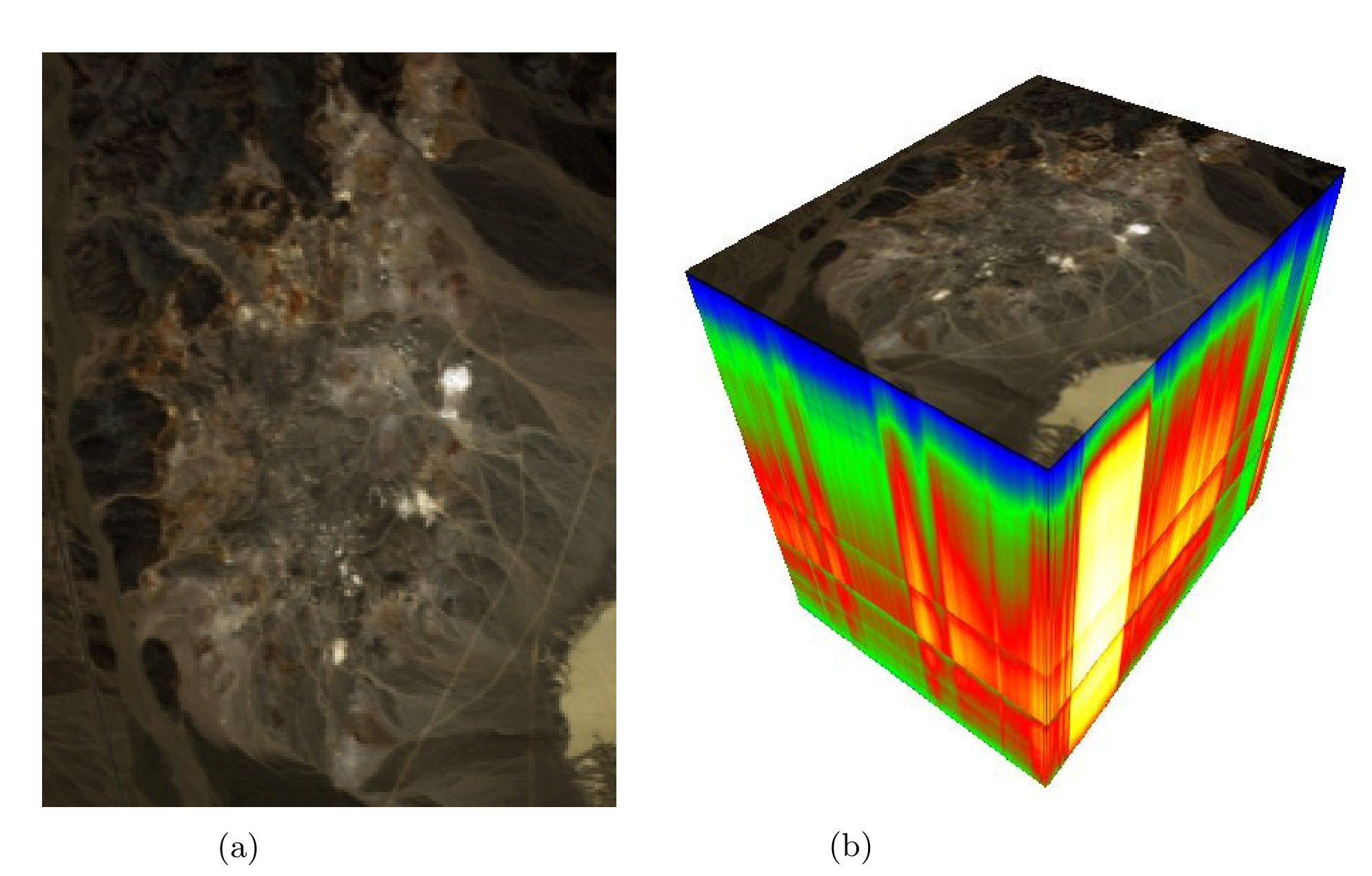}
	
	\caption{Hyperspectral data illustration of AVIRIS Cuprite dataset (a) RGB image using (40, 20, 10) bands (b) Hypercube using 188 bands.} \label{fig:hyper}
\end{figure}

Today, hyperspectral imaging has been transformed from sparse research into  a promising tool. The spectral information produced by advanced hyperspectral imaging instruments has given rise to new insights in a variety of applications such as: monitoring urban and environmental processes, preventing and observing destructive factors including weather detection, environmental hazard detection, oil spill monitoring and other types of chemical contamination \cite{XU201754}. Advanced hyperspectral instruments such as NASA’s Airborne Visible Infrared Imaging Spectrometer (AVIRIS) are able to cover the wavelengths from 400 to 2500 nm containing more than 200 spectral channels with spectral resolution of less than 10 nm \cite{henrot2016dynamical}. Each pixel vector in the resulting stack of images represents a spectral signature that characterizes the underlying materials within that limited area \cite{iordache2011sparse}. Figure~\ref{fig:hyper}.(a) shows RGB image of AVIRIS Cuprite dataset using three bands (40, 20, 10) and Figure~\ref{fig:hyper}.(b) illustrates 3D hypercube of this dataset using 188 bands after removing noisy and water absorption channels. The figures are generated using Spectral Python (SPy)\footnote{www.spectralpython.net} module.

There are different assumptions regarding the type of spectral mixing. Based on these assumptions, different mixing models have been proposed, which are divided into linear and nonlinear models:

Linear mixing model (LMM): The measured spectra are expressed as linear (convex) combination of all the spectral signatures of materials present in the mixed pixel.

Nonlinear mixing model: In this model, the spectrum recorded in each pixel is a nonlinear function of the spectrum of the principal elements, that the reflections recorded by the sensor is the result of multiple reflections and multiple interactions between materials \cite{altmann2015robust}.

Since linear mixing model is rather simple and gives prominent results, it is widely used in the literature. First category of spectral unmixing methods based on LMM assumes that pure pixels exist in the image for each of the principal elements called endmembers, so that the recorded signal of that pixel contains only the frequency spectrum of that element. Vertex component analysis (VCA) method works on this basis \cite{miao2007endmember}. Nonnegative matrix factorization (NMF) methods are another category of spectral unmixing algorithms. These methods have been used extensively to solve the problem of Blind source separation (BSS). To improve the performance of nonnegative matrix factorization methods, the sparsity constraint has been added to their cost function, since the number of nonzero elements of the abundance matrix is very low and this assumption is close to reality \cite{li2018nonnegative,wodecki2019novel}.

To estimate fractional abundance and endmembers matrices, that are blind problems, the nonnegative matrix fractional NMF and its generalized methods are particularly useful. In multilayer nonnegative matrix factorization (MLNMF) method, deep multilayer decomposition is used and fractional abundance matrices are modeled as a sparse matrix in each layer \cite{rajabi2014spectral}. Graph regularized NMF (GNMF) is an unmixing method in which sparseness constraint is considered while the geometrical structure of the hyperspectral data is preserved \cite{rajabi2015sparsity}. Another NMF method is based on adding a sparse constraint which in terms of $l_q$ norm. In this method, a network consisting of single node clusters is used, so that each hyperspectral pixel is considered as a node where the sparsity constraint expressed with diffusion least mean p-power (LMP) strategy is optimized \cite{khoshsokhan2019sparsity,wen2013diffusion}. In the clustered multitask NMF algorithm the nodes in the network are clustered by fuzzy c-means method and diffusion LMP strategy has been used to define the cost function \cite{khoshsokhan2019clustered}. 

Spectral-spatial constrained sparse unmixing (SSCSUn) describes a method in which a total variation (TV) regulator is used as a spatial weight factor to more effectively utilize spatial information \cite{xu2016spectral}. An innovative linear method that is based on $l_1-l_2$ sparsity and TV regularization has been developed to ameliorate the accuracy of hyperspectral unmixing. First, the enhancement of the sparsity based on the $l_1-l_2$ norm is investigated to show the sparsity properties of the relative fractional abundance in a sparse regression model, because the $l_1-l_2$ norm has stronger sparsity over the $l_1$ norm. Then, considering the spatial correlation between the neighboring pixels, total variation minimizes the spatial smoothness constraint. Finally, the alternating direction method of multipliers (ADMM) is used to solve the model \cite{sun2018hyperspectral}. To further improve the accuracy of spectral unmixing in the high-coherence spectral library, an algorithm based on a kernel sparse representation model with a total variation constraint is designed. To increase the effect of similarity between measurements, library atoms and hyperspectral data were added to a kernel space in which sparse regression algorithms are used \cite{weng2020kernel}. In semi-supervised spectral unmixing model, it is assumed that the spectral signature of the elements can be obtained from the hyperspectral library of the observed image. This indicates the problem of sparse regression. Sparse unmixing via variable splitting augmented Lagrangian (SUnSAL) method can be used to solve this problem \cite{bioucas2010alternating}. The goal of iterative spectral mixture analysis (ISMA) algorithm is to find the optimal endmember set to estimate abundance fractions \cite{rogge2006iterative}. On the other hand, subspace matching pursuit (SMP) is one of the sparse methods used to solve the problem of blind sources separation \cite{shi2013subspace}. Robust spectral unmixing (RSU) algorithm uses the mean squares error to solve the problem of spectral unmixing \cite{ma2016robust}. In contrast with the above-mentioned methods where the average value of Gaussian noise used for all spectral bands, the sparse unmixing method with the bandwise unmixing model (SUBM) uses the idea of Gaussian noise difference in each band for spectral separation of hyperspectral images \cite{li2020sparse}. In the spatial discontinuity-weighted (SDW) sparse unmixing method, the objective is to use spatial information on the edges of the hyperspectral images. This information, extracted using Sobel filter, is then weighted for sparse unmixing \cite{zhang2018spatial}.

Recently, deep learning based methods are introduces to solve hyperspectral unmixing problem. A general deep learning framework for self-supervised hyperspectral unmixing called EGU-Net (Endmember-Guided Unmixing Network) is introduced by Hong et. al. \cite{EGU_Net22}. EGU-Net uses a two-stream Siamese network to learn a network from pure or nearly-pure endmembers for the purpose of correcting the weights of another network by adding spectral unmixing constraints like abundance sum to one and positivity. Another deep learning method for hyperspectral unmixing is proposed by Vijayashekhar et. al. \cite{DeepGRSL22} based on two-stage fully connected self-supervised network. In this method the main goal is reconstructing the hyperspectral data using a jointly optimized network based on two-stage loss function.

In this paper, sparse Bayesian strategy and block sparse structure \cite{korki2016, korki2016_recovery} are used for spectral unmixing of hyperspectral images. The basis of the Bayesian learning strategy is used on hyperparameter information attributed to each pixel. In the proposed method, block sparse structure of the hyperspectral images is exploited. In this way, in addition to using the hyperparameter information attributed to each pixel, hyperparameters of the neighbours of that pixel are also used, which improves Bayesian learning performance.

The rest of this paper is organized as follows. In the second section, mathematical expression of the problem is discussed. In the third section, the sparse signal structure which is the basis of our proposed method, is explained. In the fourth section, we described the proposed algorithm using the sparse Bayesian learning (SBL) strategy. In the fifth section, for the performance evaluation, the proposed algorithm is evaluated using synthetic and real dataset. Proposed method is compared with three competing methods and the results are presented. Finally in the conclusion section, our concluding remarks are presented.

\section{Problem Formulation}

In the linear mixture model, it is assumed that the spectral response of each pixel in each spectral band is a linear combination of all the end members present in the pixel. For each pixel, the linear mixing model equation is as follows:

\begin{equation}
	y_i=\sum_{j=1}^{q}{a_{ij}x_j+n_i}
	\label{eq:lmm}
\end{equation}

Where $y_i$ is the reflectance value at $i$th spectral band, $a_{ij}$ is the reflectance value of the $j$th endmember at the $i$th spectral band, $x_j$ is the fractional abundance of the $j$th endmember, $n_i$ represents the measurement noise for the spectral band $i$, and $q$ is the total number of endmembers. If it is assumed that the hyperspectral sensor collects data in L-band, (1) can be written in matrix form:

\begin{equation}
	\mathbf{y}=\mathbf{Ax} + \mathbf{n}
\end{equation}

Where $\mathbf{y}$ is an $L\times1$ vector, corresponding to the measured spectrum of the pixel, $\mathbf{A}$ is an $L\times q$ matrix comprising of $q$ pure spectral signatures (endmembers), $\mathbf{x}$ is a $q\times1$ vector containing the fractional abundances of the endmembers, and $\mathbf{n}$ is an $L\times1$ vector representing the measurement noise affecting each spectral band. The number of endmembers $q$ is assumed to be known in our experiments. To estimate $q$, virtual dimensionality estimation methods for hyperspectral data can be used \cite{VirtualDim_JSTARS_2018, JSTARS20_MultipleHypo}. The abundance vector of the pure elements in the combination of each pixel is usually associated with the following two constraint:

\begin{equation}
	\mathbf{x}\geq0
\end{equation}
\begin{equation}
	\mathbf{1}^{T}\mathbf{x}=1
\end{equation}

where $1^T$ is a row vector of 1’s. These equations are abundance nonnegativity constraint (ANC) and abundance sum to one constraint (ASC) respectively.

\section{Proposed Methodology}
In this section, our proposed methodology based on block sparse Bayesian learning is presented. 

\subsection{BLOCK SPARSE SIGNAL REPRESENTATION}

The Shannon-Nyquist theory states that for a complete reconstruction of a bounded signal, the sampling rate must be at least twice the maximum frequency in the signal; whereas, in compressed sensing, less number of samples or measurements is required \cite{farrow2011nyquist}. Compressed sensing or sampling is based on the principle that if a signal in a base or dictionary has a sparse representation, the signal can then be recovered with a far less number of measurements compared with the signal length \cite{wen2019sharp}. 

In many practical issues, zero coefficients appear as cluster, which is called a block sparse structure. A number of sparse signal recovery methods utilize this structure like the pattern-coupled sparse Bayesian learning (PCSBL) method \cite{PCSBL2015} that is described in Section 4. Hyperspectral images have a block sparse structure due to the proximity of similar materials. In this paper, a new algorithm for spectral unmixing based on block sparse recovery methods is presented in Section 4.

\subsection{SPARSE BEYESIAN LEARNING}

Here, we address the problem of block sparse signal recovery of vector measurement views that is expressed in Equation~\ref{eq:lmm}. The abundance fractions vector has a block sparse structure, but the exact properties of the block such as the location and size of each block are not known. In the sparse Bayesian learning framework, it is modeled as a Gaussian prior distribution:

\begin{equation}
	p(x\vert \alpha)=\prod_{i=1}^n{p(x_i\vert \alpha_i)}
	\label{eq:pxalpha}
\end{equation}

Where $\alpha\overset{\Delta}{=}{\{\alpha_i\}}$ are non-negative hyperparameters controlling the sparsity of the signal x. From Equation~\ref{eq:pxalpha}, it is understood that when $\alpha_i$  approaches infinity, the corresponding coefficient $x_i$ becomes zero. The hyperparameters ${\alpha_i}$ can be learned by maximizing their posterior probability. This property of the coefficients provides the possibility of retrieving block sparse signals in a more reliable manner. In sparse Bayesian learning model each parameter is dependent on its neighbors in addition to the coefficient itself. Therefore, if one of the coefficients is identified, this identification will affect the improvement of its two neighboring coefficients \cite{wang2018alternative}. In this manner, by considering the neighbors, a framework for describing block sparse signals will be provided. The sparse Bayesian learning use Gamma distributions as hyperpriors over the hyperparameters.

\subsection{Proposed Bayesian algorithm with known noise variance}

The sparse Bayesian learning method can be used in recovery of block sparse signals. For ease of exposition, we first assume that the noise variance $\sigma^2$ is known a priori. Therefore, based on sparse Bayesian learning, the posterior distribution of x is expressed as follows:
\begin{equation}
	p(x\vert \alpha,y) \propto p(x\vert \alpha)p(y\vert \alpha)
\end{equation}
It can be verified that the posterior probability $p(x\vert \alpha,y)$ has a Gaussian distribution with the following mean and covariance
\begin{equation}
	\mu = \sigma^2\phi A^{T}y
\end{equation}
\begin{equation}
	\phi = (\sigma^2A^TA+D)^{-1}
\end{equation}
Where D is a diagonal matrix with its $i$th diagonal element equal to \cite{PCSBL2015}:
\begin{equation}
	D_i \overset{\Delta}{=} (\alpha_i+\beta \alpha_{i-1} + \beta \alpha_{i+1})
\end{equation}
$\beta$ is a parameter indicating the pattern relevance between the coefficient and its neighboring coefficients $x_{i+1},x_{i-1}$. When $\beta=0$, the prior distribution reduces to the prior for the conventional sparse Bayesian learning. When $\beta>0$, we see that the sparsity of $x_i$ is not only controlled by the hyperparameter $\alpha_i$, but also by the neighboring hyperparameters $\alpha_{i+1},\alpha_{i-1}$. Given a set of estimated hyperparameters $\alpha_i$, the maximum a posterior (MAP) estimate of x is the mean of its posterior distribution,
\begin{equation}
	x_{MAP} = \mu = (\sigma^2 D+A^T A)^{-1} A^T	
\end{equation}
with respect to $x$, the ${\alpha_i}$ hyperparameters using the Maximum Expectation (EM) algorithm are estimated as follows \cite{PCSBL2015}:
\begin{equation}
	\hat{\alpha_i} = k/(0.5\omega_i+10^{-4})\;\;\;\;\forall i=1,…,n
\end{equation}
where $k>0$ and $\omega_i$ represents the mean and weighted covariance as follows:
\begin{equation}
	\omega_i=(\hat{\mu}_i^2+\hat{\phi}_{i,i})+(\hat{\mu}_{i+1}^2+\hat{\phi}_{i+1,i+1})+(\hat{\mu}_{i-1}^2+\hat{\phi}_{i-1,i-1})
\end{equation}
Updating hyperparamter $\alpha_i$ as well as x continues until the stopping criteria in Equation~\ref{eq:stopping} is achieved \cite{fessler1994space}.

\begin{equation}
		\vert\vert R \vert\vert_2 = \vert\vert \mu_{new}-\mu_{old} \vert\vert_2 \leq \epsilon
	\label{eq:stopping}
\end{equation}
where $\mu_{new}$ and $\mu_{old}$ are estimated values of abundance vectors $x$ in two consecutive iterations. In this paper $\epsilon$ is set to $10^{-8}$.

\subsection{Proposed Bayesian algorithm with unknown noise variance}
In the previous section it was assumed that the noise variance is known. In this section, as is generally the case in reality, it is assumed that the noise variance is unknown. In this case, a new hyperparameter named $\gamma$ is introduced that represents the variance noise:
\begin{equation}
	p(\gamma)= \Gamma(c)^{-1}d^c\gamma^c e^{-d\gamma}
\end{equation}
where $c=d=10^{-4}$. The only difference between this method and the one in the previous section is the estimation of the hyperparameters $\alpha$ and the variance of the noise (or equivalent $\gamma$) that is again used in the EM algorithm. After applying the EM algorithm, the $\alpha$ and $\gamma$ estimations are performed, where the estimate of $\alpha$ is the same as in the previous section. The only difference in this method is the noise equivalent estimation, $\gamma$, that is calculated as follows:
\begin{equation}
	\frac{1}{\gamma^{t+1}} = \frac{\vert\vert y-Ax \vert\vert_2^2+(\gamma(t))^{-1}\sum_{i=1}^{n}\rho_i+2d}{m+2c}
\end{equation}
Where $\rho_i \overset{\Delta}{=} 1-\hat{\phi}_{i,i}(\alpha_i^{(t)}+\beta \alpha_{i-1}^{(t)}+\beta \alpha_{i+1}^{(t)})$ and $m$ indicates the number of spectral bands. The update of $\gamma$ is similar to the update procedure described in the previous section, until the stopping criterion stated in Equation~\ref{eq:stopping} is met.

However, the assumption of known noise variance is not critical and the noise variance can be estimated from the measurements suggested by \cite{Bayesian09} in the form of:

\begin{equation}
	\hat\sigma^2=||y-A\hat{x}||^2/N 
\end{equation}

\begin{figure}[H]
	\centering
	\includegraphics[width=5cm]{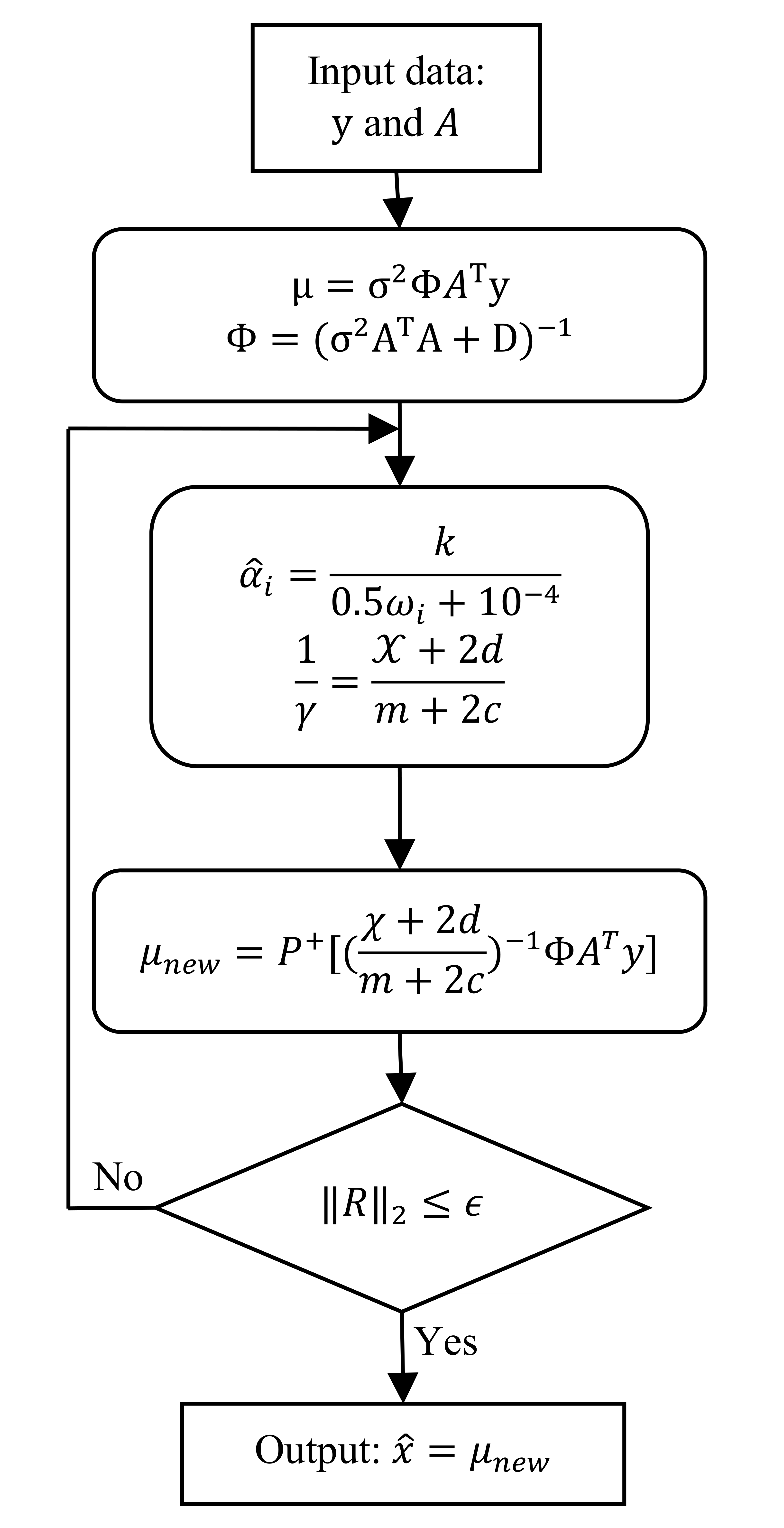}
	\caption{Flow chart of the proposed algorithm.\label{fig:FlowChart}}
\end{figure}

The overall flowchart of the proposed algorithm is depicted in Figure~\ref{fig:FlowChart}. Details and equations in this flowchart are discussed in the previous subsections.

\section{Experiments and Results}
Two types of data have been used to conduct experiments for the performance evaluation of the proposed algorithm: synthetic and real datasets. The real dataset is recorded using hyperspectral sensors and is publicly available, and the synthetic dataset is generated using spectral libraries.

\subsection{Synthetic Data}
Synthetic images are generated using a set of spectral signatures selected from publicly available spectral libraries, such as the USGS Digital Spectral Library \cite{USGS07_splib06}. The data consists of 224 spectral bands, covering wavelengths ranging from 0.38 to 2.5 $\mu m$ with a spectral resolution of 10 nm. To make linear combinations, the whole image is divided into small 5x5 blocks, the pixels within each block are pure and so have the same spectral signature, randomly selected from a set of endmembers. To simulate an image with mixed pixels, the resulting image passes through a low pass filter. In this way a hyperspectral scene in which mixed pixels exist were generated. The used low pass filter is a simple averaging filter, $k\times k$, where $k$ is controlled by the degree of mixing. To remove pure pixels and generate highly mixed data, all pixels with fractional abundance greater than eighty percent were replaced with a mixture of all pixels with the same fractional abundance. That means a mixture in which each endmember has abundance of $1/c$, where $c$ is the number of endmembers. To simulate sensor errors and noises, Gaussian noise with zero mean is added to the combination, and it is assumed that the noise is spatially and spectrally uncorrelated. Therefore, the covariance matrix of noise will be equal to  $\sigma^2 I$. The signal-to-noise ratio (SNR) can be calculated using $\text{SNR}=10log10 (\mathbb{E}[x^T x]/\mathbb{E}[n^T n])$, where $\mathbb{E}[.]$ is an expectation operator.

To compare the results quantitatively, abundance angle distance (AAD) and mean square error (MSE) evaluation criteria are used. These are common criteria in hyperspectral unmixing studies that are used to measure the similarity of results and reference values. The AAD criterion is defined as follows:
\begin{equation}
	AAD = cos^{-1}(\frac{x^T \hat{x}}{||x|| ||\hat{x}||})
\end{equation}
Where $x$ is the actual abundance vector of a pixel and $\hat{x}$ is the estimated abundance vector.
Another evaluation criterion is the MSE, which is calculated based on difference between the recorded and estimated results.
\begin{equation}
	MSE = \frac{1}{n} \sum_{i=1}^{n}(y_i-\hat{y}_i)^2
\end{equation}
where $\hat{y}_i$  and  $y_i$ are the estimated spectrum and the measured spectrum received by sensors for $i$th pixel, respectively.

The proposed algorithm is first implemented on synthetic data. The parameters to generate this dataset are set as follows. Twelve spectral signatures from the USGS library were randomly selected. A $5\times 5$ low pass filter is used. Then Gaussian noise with zero mean at 6 different SNR values, i.e. SNR=[15,20,25,30,35,40] is added to the generated data. For the different values of the parameter $\beta = [0.1,0.5,1]$ the proposed algorithm is implemented. 

\begin{figure}[H]
	\centering
	\includegraphics[width=10cm]{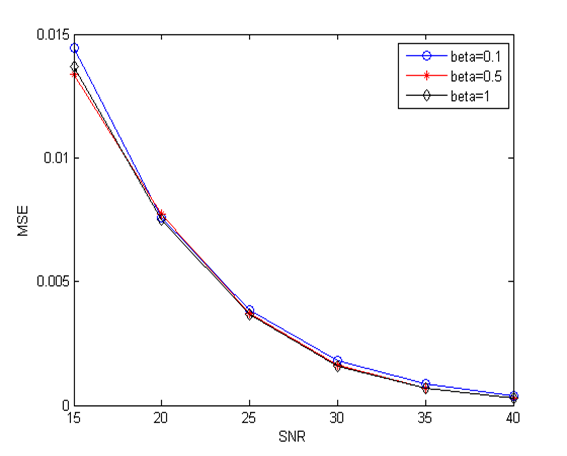}
	\caption{Mean Square Error (MSE) relative to different $\beta$s in different SNR ranges in the proposed algorithm.\label{fig:MSE_SNR}}
\end{figure}

\begin{figure}[H]
	\centering
	\includegraphics[width=10cm]{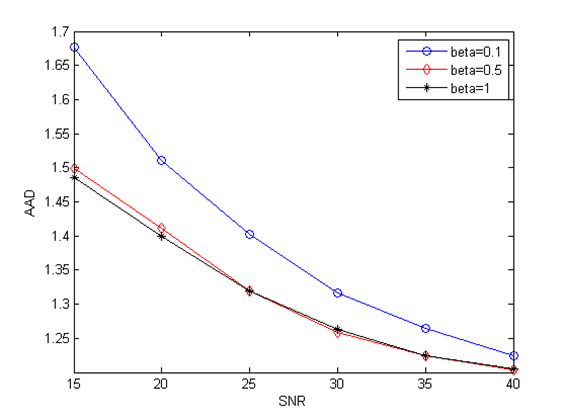}
	\caption{Abundance Angle Distance (AAD) relative to different $\beta$s in different SNR ranges in the proposed algorithm.\label{fig:AAD_SNR}}
\end{figure}

First, based on the $\beta$ parameter, the evaluation charts are plotted according to AAD and MSE criteria. As it can be seen from both Figures~\ref{fig:MSE_SNR} and \ref{fig:AAD_SNR}, when the value of $\beta$ increases, both the MSE and AAD criteria decrease, indicating the convergence of the proposed algorithm to the least possible error. With increasing SNR, MSE and AAD criteria decrease, which is due to the high level of SNR value. The results is not sensitive to the value of $\beta$ as it can be seen in Figure~\ref{fig:MSE_SNR}. Selection of a nonzero value of $0<\beta<1$ is sufficient for recovery improvement.

\begin{figure}[H]
	\centering
	\includegraphics[width=10cm]{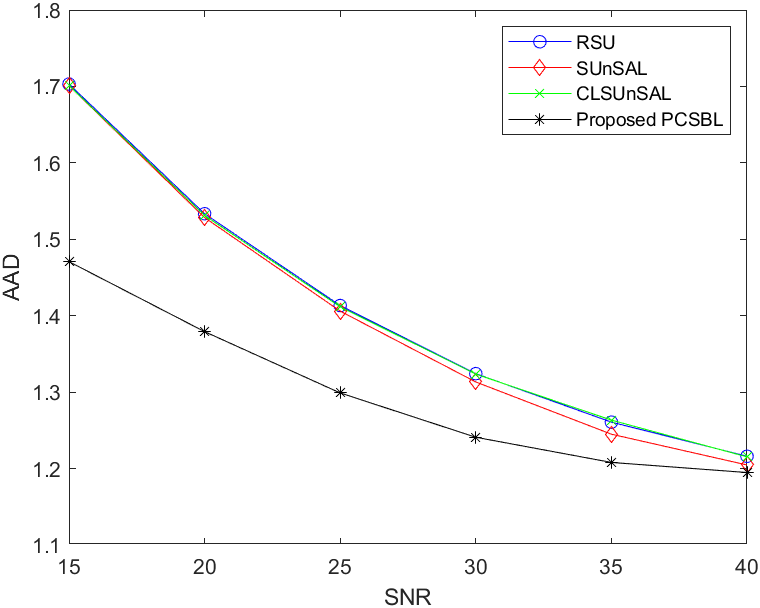}
	\caption{Comparison of Abundance Angle Distance(AAD) of Proposed Algorithm with SUnSAL and RSU Algorithms.}\label{fig:MSE_Compare}
\end{figure}

\begin{figure}[H]
	\centering
	\includegraphics[width=10cm]{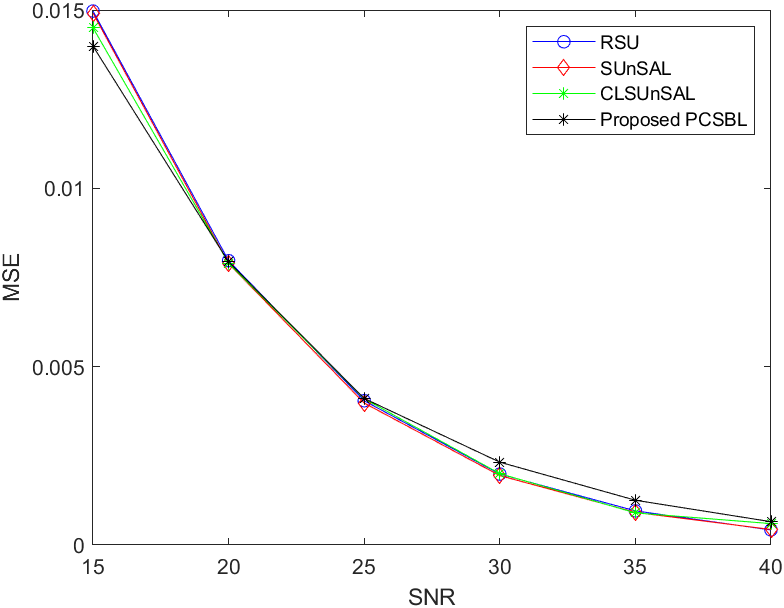}
	\caption{Comparison of Mean Squared Error (MSE) of Proposed Algorithm with SUnSAL and RSU Algorithms.}\label{fig:AAD_Compare}
\end{figure}

In Figures~\ref{fig:MSE_Compare} and \ref{fig:AAD_Compare}, the results of the proposed algorithm is compared with SUnSAL \cite{bioucas2010alternating} and RSU \cite{ma2016robust} algorithms in different SNRs and based on AAD and MSE evaluation criterion.

\begin{figure}[H]
	\small
	\centering
	\includegraphics[width=17cm]{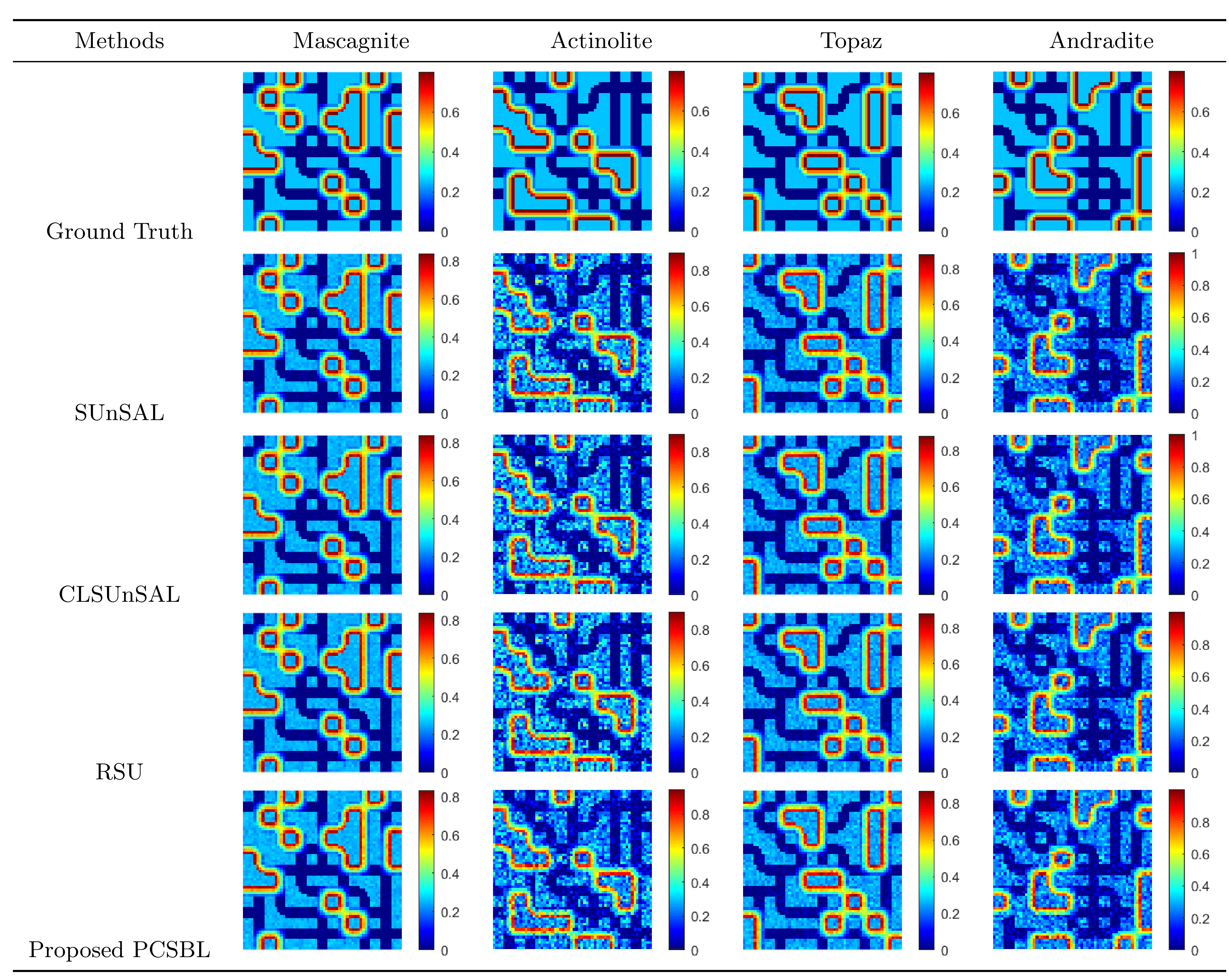}
	\caption{Fractional abundance map of endmembers for synthetic dataset at SNR=15dB.}\label{fig:synth}
\end{figure}

Tables~\ref{tab:table1} and ~\ref{tab:table2} show the MSE and AAD metrics in different SNRs, respectively. In Figure~\ref{fig:synth}, the estimated fractional abundance map of the two synthetic endmembers at SNR=25dB resulted from the proposed algorithm is compared with SUnSAL and RSU algorithms' results. The comparison indicates that the proposed method more accurately estimates the fractional abundance map in comparison to the two other competing methods.

\begin{table}[h]
	\begin{center}
		\caption{Comparison of different methods in terms of MSE for different SNRs}
		\label{tab:table1}
		\begin{tabular}{l|c|c|c|c|c|c} 
			\textbf{SNR} & \textbf{15} & \textbf{20} & \textbf{25} & \textbf{30} & \textbf{35} & \textbf{40}\\
			\hline
			SUnSAL \cite{bioucas2010alternating} & 0.01491 & 0.00789 & 0.00400 & 0.00194 & 0.00091 & 0.00043\\
			\hline
			CLSUnSAL \cite{CLSUnSAL15} & 0.01489 & 0.00785 & 0.00396 & 0.00192 & 0.00089 & 0.00044\\
			\hline
			RSU \cite{ma2016robust} &0.01497 &	0.00797	& 0.00404 &	0.00198	& 0.00095 &	0.00043\\
			\hline
			Proposed PCSBL & \textbf{0.01398} &	\textbf{0.00793} &	\textbf{0.00396}	& \textbf{0.00190} & \textbf{0.00080} &	\textbf{0.00040} \\
		\end{tabular}
	\end{center}
\end{table}

\begin{table}[h]
	\begin{center}
		\caption{Comparison of different methods in terms of AAD for different SNRs}
		\label{tab:table2}
		\begin{tabular}{l|c|c|c|c|c|c} 
			\textbf{SNR} & \textbf{15} & \textbf{20} & \textbf{25} & \textbf{30} & \textbf{35} & \textbf{40}\\
			\hline
			SUnSAL \cite{bioucas2010alternating}  & 1.701 & 1.528 & 1.405 & 1.405 & 1.313 & 1.244\\
			\hline
			CLSUnSAL \cite{CLSUnSAL15}  & 1.702 & 1.485 & 1.463 & 1.389 & 1.297 & 1.240\\
			\hline
			RSU \cite{ma2016robust} & 1.695 & 1.533 & 1.413 & 1.324	& 1.260 & 1.216 \\
			\hline
			Proposed PCSBL & \textbf{1.470} &	\textbf{1.379} &	\textbf{1.298}	& \textbf{1.240} & \textbf{1.208} &	\textbf{1.194} \\
		\end{tabular}
	\end{center}
\end{table}

\subsection{Real Data}
\begin{figure}
	\centering
	\includegraphics[width=17cm]{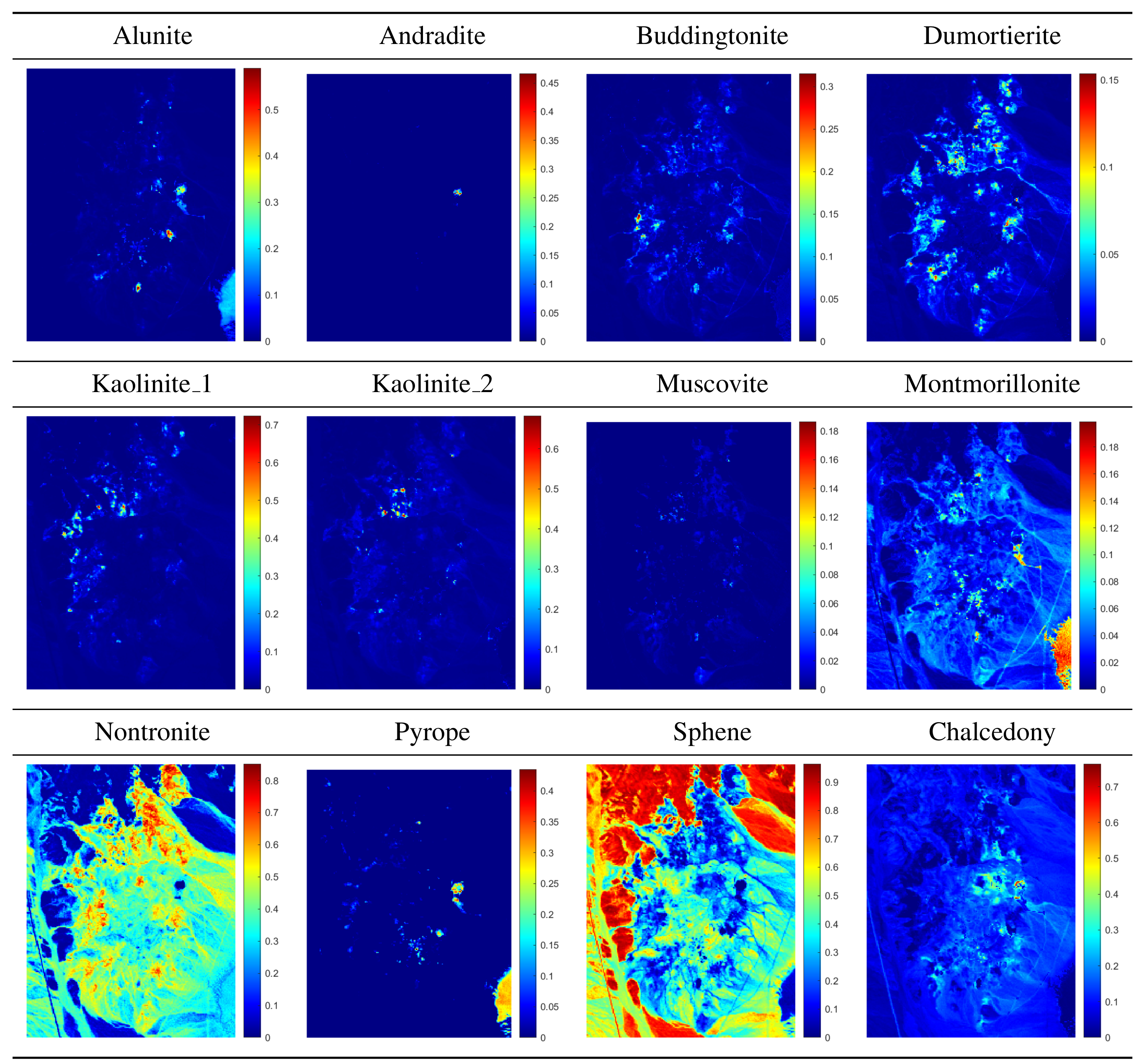}
	\caption{Fractional abundance map of all 12 endmembers of Cuprite real dataset using proposed PCSBL method.\label{fig:real}}
\end{figure}

To date, many hyperspectral sensors have recorded images in different missions that are used to test hyperspectral data processing algorithms. One of these hyperspectral sensors is NASA's AVIRIS sensor that is introduced in Section~\ref{sec:intro}. The data we have used in this research is from Cuprite region in Nevada \cite{AVIRIS_Cuprite}. There are many benefits using this data, including that this place has been used for remote sensing experiments since 1980 and a lot of high precision research is available on this area. To improve the separation performance, 36 low-SNR bands, that are significantly distorted with noise due to atmospheric effects, have been excluded from the 224-bands data cube. The remaining 188 bands have been then used in the experiments. In Figure~\ref{fig:real}, the estimated fractional abundance maps of the proposed algorithm are compared with thos of SUnSAL and RSU methods for two endmembers of the Cuprite data. Results show that the proposed method can effectively estimate the abundance fractions maps.

\section{Conclusion}

In this paper, a new method for spectral unmixing of hyperspectral images is presented that uses sparse Bayesian learning strategy for recovery of block sparse signals whose structure is unknown. In the Sparse Bayesian learning model, the sparsity of each coefficient depends only on the relevant hyperparameter. But in the proposed method, the sparsity also depends on its neighborhood hyperparameters. By maximizing posterior probability, hyperparameters in sparse signals can be estimated based on an iterative algorithm using expectation maximization method. To evaluate the performance of the proposed algorithm, both synthetic data and real AVIRIS Cuprite dataset are used. Results based on quantitative criteria MSE and AAD show that the proposed algorithm using the block sparse structure has better performance compared to other methods, even without knowing the exact location and size of each block. For future research, block sparse methods such as Block-IBA that do not need to know the block structure of the signal can be used for hyperspectral unmixing.


\bibliography{refs}   
\bibliographystyle{spiejour}   


%

\listoffigures
\listoftables

\end{spacing}
\end{document}